\documentstyle[12pt,epsf]{article}
\newcommand{\text}[1]{\hbox{\rm \ #1\ \/}}

\newcommand{\CC}{\mbox{${\rm \:  C\!\!\! I
\;\;}$}}

\newcommand{\RR}{\mbox{${\rm \:  R\!\!\!\! I
\;\;}$}} 

\newcommand{\qed}{\hfill $\Box$ \vskip 2ex}
\newcommand{\be}[1]{\begin{equation}\label{#1}}
\newcommand{\ee}{\end{equation}}
\newcommand{\vs}{\vspace{0.25cm}}

\begin{document}

\title{Optimal evaluation of generalized 
Euler angles with applications to classical and quantum control}

\author{Domenico D'Alessandro\\ 
Department of Mathematics\\
Iowa State University \\
Ames, IA 50011,  USA\\
Tel. (+1) 515 294 8130\\
email: daless@iastate.edu}

\maketitle

\begin{abstract}

Given two linearly independent 
matrices in $so(3)$, $Z_1$ and 
$Z_2$, every rotation matrix $X_f 
\in SO(3)$ can be written as  the 
product of alternate elements from 
the one dimensional subgroups corresponding 
to $Z_1$ and $Z_2$, namely $X_f=e^{Z_1 t_1}e^{Z_2 t_2}e^{Z_1t_3} \cdot
\cdot \cdot e^{Z_1t_s}$. The parameters $t_i$, $i=1,...,s$ are
called {\it generalized Euler angles}.

In this paper, we evaluate the minimum number of factors required for
the  factorization of $X_f \in SO(3)$, as a function of $X_f$,  
and provide an algorithm to determine
the  generalized Euler angles explicitly. The results can be applied to
the bang bang control with minimum number of switches of some classical
control systems and of two level quantum systems.    

\end{abstract}

\vspace{0.25cm}


\section{Introduction}
In this paper, we deal with the problem of steering control for bilinear
 systems of the form 
\be{EU71}
\dot x=Ax+Bxu, 
\ee
where $x \in \RR^3$, $u$ is a control function, 
and $A$ and $B$ are skew-symmetric $3 \times 3$ matrices,
namely matrices in $so(3)$. Several systems in applications  have the
structure (\ref{EU71}). In particular the most common  example is
given by the dynamics of the rigid body \cite{greenwood} where one
component of the angular velocity  is seen as the  control $u$ and the
others are held constant. The fundamental matrix of equation
(\ref{EU71}) represents the orientation  of the rigid body. 
Another example is the lossless electrical network dealt with in 
\cite{RamakCDC}. A two level quantum system driven by a single time
varying component of an electro-magnetic field also has the structure
(\ref{EU71}) \cite{conmohIEEE}, where $x$ represents the 
state $\in \CC^2$ and the matrices
$A$ and $B$ are in the Lie algebra $su(2)$. Because of the connection
between the Lie groups $SO(3)$ and $SU(2)$ the results presented here
can be applied to the latter system as well.

The fundamental matrix of the system
(\ref{EU71}), $X$, satisfies 
\be{EU72}
\dot X=AX+BXu, 
\ee  
with initial condition equal to the $3 \times 3$ identity matrix. It
follows from the results of \cite{suss} that if $A$ and $B$ are linearly
independent (and therefore generate $so(3)$ which has dimension $3$ and
no two-dimensional subalgebras) a 
piecewise constant control is sufficient to steer the state of
(\ref{EU72}) from the identity to every matrix $X_f$ in $SO(3)$ and, as a
consequence, the state $x$ of (\ref{EU71}) between two states with equal
length. Let us assume now that the control $u$ is allowed to attain only two
values, $M$ and $N$. Define $Z_1:=A+BM$ and $Z_2:=A+BN$, and assume a
factorization of the desired target state $X_f$ of the type 
\be{EU73}
X_f=e^{Z_1 t_1}e^{Z_2 t_2}e^{Z_1t_3} \cdot
\cdot \cdot e^{Z_1t_s}, 
\ee
is known with $t_1,t_2,...,t_s >0$\footnote{This is done without loss of
generality since the one parameter subgroups corresponding to $Z_1$ and
$Z_2$ are closed, namely the functions $e^{Z_{1,2}t}$ are
periodic.}. Then a piecewise constant control equal to $M$ for time
$t_s$, $N$ for time $t_{s-1}$, $M$ for time $t_{s-2}$ and so on, 
drives the state of (\ref{EU72}) from the identity to $X_f$ in
(\ref{EU73}). This idea, involving Lie group decompositions,  
has  recently been used to prescribe controls for
quantum mechanical systems where the underlying Lie group is the group
of special unitary matrices of dimension $n$, $SU(n)$ (see
e.g. \cite{MIKOCDC}, \cite{HOMO}, 
\cite{Khaneja}, \cite{RAMA1}, \cite{Sonia1} and
references therein). If the control is bounded in magnitude, 
namely $|u| \leq M$ we can choose $N:=-M$ and $M$ as the two values for
the control. From a practical point of view one would like to have a
factorization of $X_f$ in terms of the matrices $Z_1$ and $Z_2$ that
involves the least number of factors, so that the control law has the
minimum number of switches. Moreover, an algorithm is needed to
evaluate  the  {\it generalized Euler angles} 
$t_i$, $i=1,...,s$. This paper is devoted to
the solution of these two problems. Constructive factorizations of
$SU(2)$ and $SO(3)$ can be found in the papers \cite{HOMO}, \cite{RAMA1}
that, however, do not consider the problem of minimizing the number of
factors.  

\vs

The paper is organized as follows. 
In the next section we give some
preliminary definitions that will 
be used in the following and recall
some results proved in \cite{lowen} 
concerning factorizations of elements 
of the Lie group $SO(3)$ of the type 
(\ref{EU73}). We also transform, using
a change of coordinates,  every
pair of linearly independent matrices 
$\in so(3)$ into  a canonical form,
that will be used in the following sections, without loss of
generality. In Section 3 we evaluate the minimum number of factors
needed in a factorization of  matrix $X_f \in SO(3)$ of the type (\ref{EU73})
given $Z_1$ and $Z_2$. In Section 4 we give an algorithm for the
determination of the generalized Euler angles $t_1,...,t_s$. 
We discuss applications to the 
control of classical and quantum systems 
in Section 5.

\section{Preliminaries} 

The  inner product $ < \cdot, \cdot>$ 
 between two elements of $so(3)$, $Z_1$ and $Z_2$ is defined as 
\be{EU74}
< Z_1,Z_2> = Trace (Z_1 Z_2^T). 
\ee
If $<Z_1,Z_2>=0$, the maximum number of factors $s$  needed to express a 
matrix $X_f$ in $SO(3)$ as in  (\ref{EU73}) (maximum over $SO(3)$) 
is three, and the factorization in
(\ref{EU73}) is the classical Euler resolution of a rotation (see
e.g. \cite{RAO}) (modulo a change of coordinates and a re-scaling of the
variables $t$). The parameters $t_i$ are called {\it Euler angles} and
their calculation is standard matter (see e.g. \cite{RAO}, pg. 297).

\vs 

In \cite{lowen},  it was shown that, 
for every pair of matrices $Z_1$, $Z_2$, the number 
of factors needed to express an
element $X_f \in SO(3)$ is uniformly bounded, over $SO(3)$ 
(see also  \cite{mikoSCL} and \cite{CSL3} for
generalizations to every compact Lie group). 
%%%%%%%%%%%%%%%%%%%%%%%%%%%%%%%%%
%%%%%%%%%%%%%%%%%%%%%%
The maximum  value
for $s$  (maximum over $SO(3)$) is called the {\it order of generation} 
of $SO(3)$ with respect to $Z_1$ 
and $Z_2$.  It has been calculated in 
\cite{lowen} and it only depends on the value of the cosine of the angle
between $Z_1$ and $Z_2$, namely 
\be{psi}
\psi:=\frac{<Z_1, Z_2>}{<Z_1,Z_1>^{\frac{1}{2}}<Z_2,
Z_2>^{\frac{1}{2}}}.  
\ee  
%the cosine of the angle between $Z_1$ and $Z_2$. If $f \geq 1$ is such
%that 
%\be{lowencond}
%cos(\frac{\pi}{f}) < |\psi| \leq cos ( \frac{\pi}{f+1}), 
%\ee
%then the order of generation is $s=f+2$. 
If $\psi=0$, the order of generation is equal to $3$ and  we
obtain the classical Euler resolution of a rotation. 
Our treatment
in the following was inspired by the proof in 
\cite{lowen}. However,   most of the treatment in \cite{lowen} is 
carried out using stereographic projections and translating the problem
to the induced subgroup of the Moebius group.  We shall treat the
factorization of  every element in $SO(3)$  by working on the 
unit sphere in $\RR^3$ and looking at $SO(3)$ as
 a transformation group on the sphere \cite{MZ}.

\vs

We now  show that there is no loss 
of generality in assuming that $Z_1$
and $Z_2$ in (\ref{EU73}) have a special form which we shall describe. 
We shall call $S_{hk}$, $h <k$,   
the matrix in $so(3)$ which has
zeros everywhere except in the $h,k$-th ($k,h$-th) 
entry which is  equal to $1$ ($-1$). 
Given a matrix $Z_1$, there exists a matrix $T_1 \in SO(3)$ such that 
\be{trans1}
T_1Z_1T_1^T=\lambda_1 S_{12},   
\ee
$\lambda_1 \not= 0$. 
This can be easily seen by choosing $T_1:=[v_1, v_2,v_3]^T$, with $v_3$
such that $v_3^T Z_1=0$ and with norm equal to one and $v_1$ and $v_2$ 
such that $\{ v_1,v_2,v_3 \}$ form an orthonormal basis 
in $\RR^3$. We also set 
\be{setting}
T_1Z_2T_1^T:=a S_{12}+ b S_{13}+c S_{23}. 
\ee 
Choose now $T_2:=e^{S_{12} \theta}$ with $\theta$ such that 
$b cos(\theta)+c sin(\theta)=0$, with $b$ and $c$ given 
in (\ref{setting}).  Then we have 
\be{another}
T_2 T_1 Z_1 T_1^T T_2^T= \lambda_1 S_{12}, 
\ee 
\be{EA1}
T_2 T_1 Z_2 T_1^T T_2^T=a S_{12} + d S_{23}, 
\ee
for some parameter $d \not= 0 $. Therefore, we can always assume that,
in appropriate coordinates, the matrices $Z_1$ and $Z_2$ have the form 
$Z_1:= \lambda_1 S_{12}$ and 
$Z_2:= aS_{12} +d S_{23}$, respectively. Moreover we
can divide $Z_1$ by $\lambda_1 \not= 0$ (this has the only effect that, in the
matrices of the form $e^{Z_1 t}$,  $t$ has to be scaled by a factor
$\lambda_1$) and analogously we can divide $Z_2$ 
(in the new coordinates in (\ref{EA1})) by $d \not= 0$ and therefore the
parameter $t$ in the subgroup $e^{Z_2t}$ has to be scaled by a factor
$d$. Define $\rho:=\frac{a}{d}$. We can assume, without loss of
generality, that the matrices $Z_1$ and $Z_2$ are given by 
\be{Z1}
Z_1:=S_{12},
\ee 
and 
\be{Z2}
Z_2:=\rho S_{12} + S_{23},
\ee 
and we shall do so in the following. Notice 
that the above manipulations do not modify the
value of the parameter 
$\psi$ in (\ref{psi}) which is given, in terms of
$\rho$, by 
\be{EA2}
\psi=\frac{\rho}{\sqrt{1+ \rho^2}}. 
\ee

\section{Decompositions with minimum number of factors} 

Assume now that an  element 
$X_f \in SO(3)$ is given, to be expressed as
in (\ref{EU73}), with $Z_1$ and $Z_2$ given in 
(\ref{Z1}), (\ref{Z2}). 
We give in this section a procedure to determine the
minimum number of factors necessary as a function of $X_f$. 

\vs

We assume $\rho$ in (\ref{Z2}) 
different from zero (the case $\rho=0$
corresponds, from (\ref{EA2}), to 
$Z_1$ and $Z_2$ orthogonal to each-other and therefore
the decomposition is the standard Euler decomposition). 
Define two sequences $\{ z_k \}$ and $\{ f_k \}$ by $z_0=f_0=-1$
\begin{eqnarray}
f_k:=\frac{1}{|\rho|} \sqrt{1-z_k^2} + z_k \label{A} \\
z_{k+1}:=\frac{2 \rho^2}{1+ \rho^2} f_k - z_k \label{B}. 
\end{eqnarray}
We have the following Lemma. 

\vs

\noindent{\bf Lemma 3.1} {\it There exists an index $\bar k \geq 1$
such that $f_k$ is defined ($|z_k| \leq 1$) for every $k \leq \bar k$,
$f_k < 1$, for every $k < \bar k$ and $f_{\bar k} \geq 1$.}

\vs 

\noindent{\bf Proof.} First notice that if $| \rho | \leq 1$, the Lemma
is true with $\bar k=1$ since $0 \leq  z_1<1$ and $f_1 \geq 1$. Let us
assume $| \rho | >1$. We first show that $f_k$ well defined and 
$f_k < 1$ implies that
$f_{k+1}$ is well defined, namely 
that $|z_{k+1}| \leq 1$.  Then we show that
there exists the first value of 
$k$,  $\bar k$, such that $f_{\bar k}
\geq 1$. 

\vs

Assume $f_k<1$. From (\ref{A}), we obtain 
\be{EUin1}
\sqrt{1-z_k^2}< | \rho | (1-z_k), 
\ee 
which gives, taking into account $|z_k| \leq 1$, 
\be{EUin2}
-1 \leq z_k < \frac{\rho ^2 -1}{1+ \rho^2}. 
\ee
Consider the expression of $z_{k+1}$ obtained 
combining (\ref{A}) and (\ref{B}), 
\be{EUin3}
z_{k+1}= \frac{2 \rho^2}{1+ \rho^2}( \frac{1}{|\rho|} \sqrt{1 -z_k^2} +
z_k)-z_k.  
\ee
Consider $z_{k+1}$ as a function of $z_k$ in 
the interval defined in (\ref{EUin2}). This
function is always increasing from the value $z_{k+1}=\frac{1 - \rho^2}{1 +
\rho^2}$ at $z_k=-1$ to the value $z_{k+1}=1$ at $z_k=\frac{\rho^2
-1}{{\rho^2+1}}$. In particular we always have $|z_{k+1}| \leq 1$
which implies that $f_{k+1}$ is well defined. To show the existence of
a $\bar k$ such that $f_{\bar k} \geq 1$, we evaluate $z_{k+1}-z_k$
using (\ref{A}) and (\ref{B}). We obtain
\be{EUin4}
z_{k+1}-z_k=\frac{2}{1+ \rho^2} ( |\rho| \sqrt{1-z_k^2}-z_k). 
\ee
Using the second inequality in (\ref{EUin2}), we obtain 
\be{EUin5}
\sqrt{1-z_k^2} > \frac{2| \rho |}{1+ \rho ^2}, 
\ee 
and plugging this into (\ref{EUin4}), we obtain 
\be{EUin6}
z_{k+1}-z_k > \frac{2}{1+ \rho^2}( \frac{2 \rho^2}{1+ \rho ^2} - z_k) > 
 \frac{2}{1+ \rho^2}, 
\ee
where, in the last inequality, we used inequality (\ref{EUin2})
again. Therefore the sequence $\{z_k\}$ is increasing by at least    
$\frac{2}{1+ \rho^2}$ at each step and since $f_k \geq z_k$ for every
$k$, from (\ref{A}), we must have a value of the index $\bar k$ such
that $f_{\bar k} \geq 1$. This concludes the proof of the Lemma. \qed 

\vs

We now relate the finite sequences $\{ z_k\}$ and $\{ f_k \}$,
$k=0,1,..., \bar k$  defined in (\ref{A}) and (\ref{B})  to the
minimum number of factors needed in the factorization
(\ref{EU73}). Consider a given target matrix $X_f := \{ x_{i,j} \}$,
$i,j=1,2,3$ to be factorized. We define a function ${\cal O}(X_f)$ which
is equal to $1$ if $x_{3,3}=-z_0=1$, it is equal to $2$ if 
$z_0 < -x_{3,3} \leq z_1$ and $x_{1,3}=\rho(-x_{3,3} +1)$ and equal to
$3$ if $z_0 < -x_{3,3} \leq z_1$ and $x_{1,3} \not = \rho( - x_{3,3}
+1)$. In cases not considered above, 
let $\tilde k$ be the highest value of 
the index $k$  such that 
\be{EUin9p}
z_{\tilde k} < -x_{3,3} 
\ee 
(recall from (\ref{EUin6}) that $z_{k}$ is increasing at each step by at
least a given amount). Then we have 
\begin{eqnarray}
{\cal O}(X_f)=2 \tilde k+2 \text{if} sign(\rho) x_{1,3} \geq - |\rho| (x_{3,3}
+f_k) \label{FORM1} \\
{\cal O}(X_f)=2 \tilde k+3 \text{if} sign(\rho) x_{1,3} <  - |\rho| (x_{3,3}
+f_k).  \label{FORM2}
\end{eqnarray} 
\vs 
The following Lemma gives the 
minimum number of factors in the 
factorization (\ref{EU73}) 
assuming that the first factor on the right
is of the form $e^{Z_1 t_s}$ with $t_s >0$. The proof of the Lemma
reveals the geometric meaning of the  finite sequences $\{ z_k \}$ and 
$\{ f_k \}$ defined in (\ref{A}), (\ref{B}). 
We denote the minimum number of factors
needed to express a general matrix 
$X_f$ as in (\ref{EU73}) by
$MIN(X_f)$. 

\vs

\noindent {\bf Lemma 3.2} {\it Assume that $X_f$ is such that the
factorization with minimum number of factors 
in (\ref{EU73}) starts with
a nontrivial factor of the type $e^{Z_1 t}$ on the right. Then
\be{minimo}
MIN(X_f)={\cal O}(X_f). 
\ee
} 

\vs

Before giving the proof of the 
Lemma, we describe the geometry of the
above construction. Considered as a transformation on the sphere of radius 
$1$ centered at
the origin,  $X_f$ transforms the South Pole 
$P_s:=[0,0,-1]^T$ into a point
$P_f:=[-x_{1,3},-x_{2,3},-x_{3,3}]$ (which is just the negative of the 
third column of $X_f$). Conversely, any matrix $\tilde X_f$ such that 
$P_f=\tilde X_f P_s$ is equal to $X_f$ up to a factor that leaves $P_s$
unchanged. Such factor will in general have the form $e^{Z_1t}$ (recall 
(\ref{Z1})) and
therefore we have $X_f=\tilde X_f e^{Z_1t}$. We would like to
find any  product with {\it minimum number} of factors  
\be{EULA11} 
\tilde X_f:= e^{Z_1 t_1}e^{Z_2 t_2}e^{Z_1t_3} \cdot
\cdot \cdot e^{Z_2t_{s-1}} e^{Z_1 t_s}, 
\ee 
(with $t_s$ possibly equal to zero) such that $P_f=\tilde X_f P_s$ and
then to obtain $X_f$ as $X_f=\tilde X_f e^{Z_1t_s}$. From the assumption
that the minimum number of factors for $X_f$ 
is obtained with a nontrivial factor  $e^{Z_1 t_s}$ on the right, the
minimum number of factors will be given by $s$. 
This observation can be interpreted in
the language of coset spaces and homogeneous spaces (see
e.g. \cite{MZ}). The subgroup ${H}:=\{
X \in SO(3)|X=e^{Z_1t}, t \in \RR \}$ is the {\it isotropy group} of the
South Pole $P_s$, namely the set of all the elements of $SO(3)$ that
leave $P_s$ fixed. There exists an 
isomorphism between elements of the
coset space  $SO(3)/H$ 
%\footnote{An element of $SO(3)/H$ is a class of
%equivalence of elements of $SO(3)$, under the 
%equivalence relation $X_1 \approx X_2 \leftrightarrow 
%X_1X_2^{-1} \in H$. They are called {\it cosets}} 
and elements  of the sphere
$S^3$. In the expression (\ref{EU73}) we use the last   term $e^{Z_1
t_s}$ to move inside a coset while the remaining factors 
 are used to go from one coset to the
other, namely from one point on the sphere $S^3$ to the other. We now
look for a transformation $\tilde X_f$ in (\ref{EULA11}) transforming
$P_s$ to $P_f$ with minimum number of factors. 

\vs

On the sphere $S^3$, every element of the form 
$e^{Z_1t}$ corresponds to a rotation about the
$z$ axis. Each point on the sphere $S^3$ 
follows a trajectory on a circle which is 
the intersection of the sphere with a horizontal
plane. The value of the $z$ coordinate 
of the point is not changed by
this rotation. Every matrix  of the form $e^{Z_2t}$ corresponds to a
rotation about the axis defined by the vector $\vec n_\rho:=[1, 0,
\rho]^T$ (points on the line through 
the origin parallel to this vector
are left invariant by the rotation). 
Under the action of this rotation,
every point on the sphere $S^3$ 
follows a trajectory on a circle which
is the intersection of a plane perpendicular to $\vec n_\rho$ and the
sphere $S^3$. Each such plane forms 
an angle $\theta:=tan^{-1} \frac{1}\rho$
with the $x-y$ plane. If we consider a trajectory $e^{Z_2t}P_s:=[x(t),
y(t), z(t)]^T$, the maximum value for the coordinate $z(t)$ will be
obtained at $z_1$ defined in (\ref{B}) (when $t=\pi$). Let us call this
point $P_1$. Following a horizontal trajectory $e^{Z_1 t} P_1$,  for
$t=\pi$, we obtain a point which is opposite to $P_1$. Let us denote
this point by $Q_1$. Following from $Q_1$ 
a trajectory  $e^{Z_2 t} Q_1$
again up to $e^{Z_2 \pi} Q1$, we obtain a point with $z$ coordinate given
by $z_2$ in (\ref{B}). The value $f_1$ is the $z-$coordinate of the
intersection of the plane 
perpendicular to $\vec n_\rho$, containing the point $ Q_1$,
 and the
$z$-axis. Notice that $z_2$ is the maximum value that can be obtained
for $z$ starting from $P_1$ and with just one switch from one type of
trajectory to the other.  Continuing this way one obtains 
the elements of the sequences 
$\{z_k \}$, $\{ f_k \}$. It follows from this geometric description that
$\{z_k\}$ is an increasing sequence and it was proven in Lemma 3.1 that
it is a {\it finite} sequence (See also the Remark following the proof
of Lemma 3.2). 
Figure 1 describes  (in a two-dimensional plot)  the  trajectories on the
sphere.  In this Figure, $\bar k$, defined in Lemma 3.1, is equal to
5. We have denoted by $P_k$, $k=0,1,...,5$ the points on the sphere with
 $z-$ coordinate equal to $z_k$, $k=0,1,...,5$. $F_k$ denotes the point
whose $z-$coordinate if $f_k$, $k=1,...,5$. 
We now use this Figure to complete the proof of the Lemma.

\vs

\noindent{\bf Proof of Lemma 3.2.} We shall refer to Figure 1 and the
above discussion. Let $P_f:=[-x_{1,3}, -x_{2,3}, -x_{3,3}]^T$. If
$x_{3,3}=1=-z_0$, then $X_f$ is of the type $e^{Z_1 t}$ and clearly
$MIN(X_f)=1$. If $z_0 < -x_{3,3} \leq z_1$ and 
$x_{1,3}=\rho (-x_{3,3} +1)$ then $P_f$ belongs to the intersection of
the plane $x+\rho(z+1)=0$ with the sphere $S^3$. The point $P_f$ can be
reached by (possibly) following a trajectory 
of the type $e^{Z_1t}$ (which leaves
$P_s:=[0,0,-1]$ unchanged) followed
by a trajectory of the type $e^{Z_2t}$. In this case, since we have
assumed that the last factor on the right in (\ref{EU73}) is  a nontrivial 
$e^{Z_1t}$ factor, we have 
$MIN(X_f)=2$. Analogously, it is easily seen that $MIN(X_f)=3$ if 
$z_0 < -x_{3,3} \leq z_1$ and $x_{1,3} \not= \rho (1 - x_{3,3})$. Now
notice that to reach a point with $z$ coordinate $\bar z$, with 
$z_{k} < \bar z \leq z_{k +1}$ we need to cross the circle 
${C}_k := \{ (x,y,z) | z= z_k, x^2+y^2+z^2=1, 
(x,y,z) \not= (x_k, y_k, z_k)  \}$. In order to cross
{\it any point} of ${C}_k$, the minimum number of pieces of trajectory 
(including possibly the first one of the type
$e^{Z_1t}$, if assumed nontrivial, and the last one to leave
${C}_k$),  is $2k+2$. This is clear when $k=1$ and follow by
induction for the other values of $k$, noticing that we must cross 
${C}_{k-1}$ before crossing ${C}_k$. To reach a point 
$P_f:=[-x_{1,3}, -x_{2,3}, -x_{3,3}]^T$ such that $z_{\tilde k} < -x_{3,3}$,
with $\tilde k \leq \bar k$, we need to cross ${C}_{\tilde k}$ and
the minimum number of factors to do that is $2 \tilde k +2$. No other
factor is needed if $P_f=[-x_{1,3}, -x_{2,3}, -x_{3,3}]^T$ is below the
plane with equation $x+ \rho (z-f_{\tilde k})=0$ 
while another factor is needed
if $P_f$ is above this plane. This 
accounts for the inequalities
(\ref{FORM1}), (\ref{FORM2}). \qed 

\vs

\begin{figure}
\epsfysize=6.50in
\epsfysize=4.50in
\centerline{\epsfbox{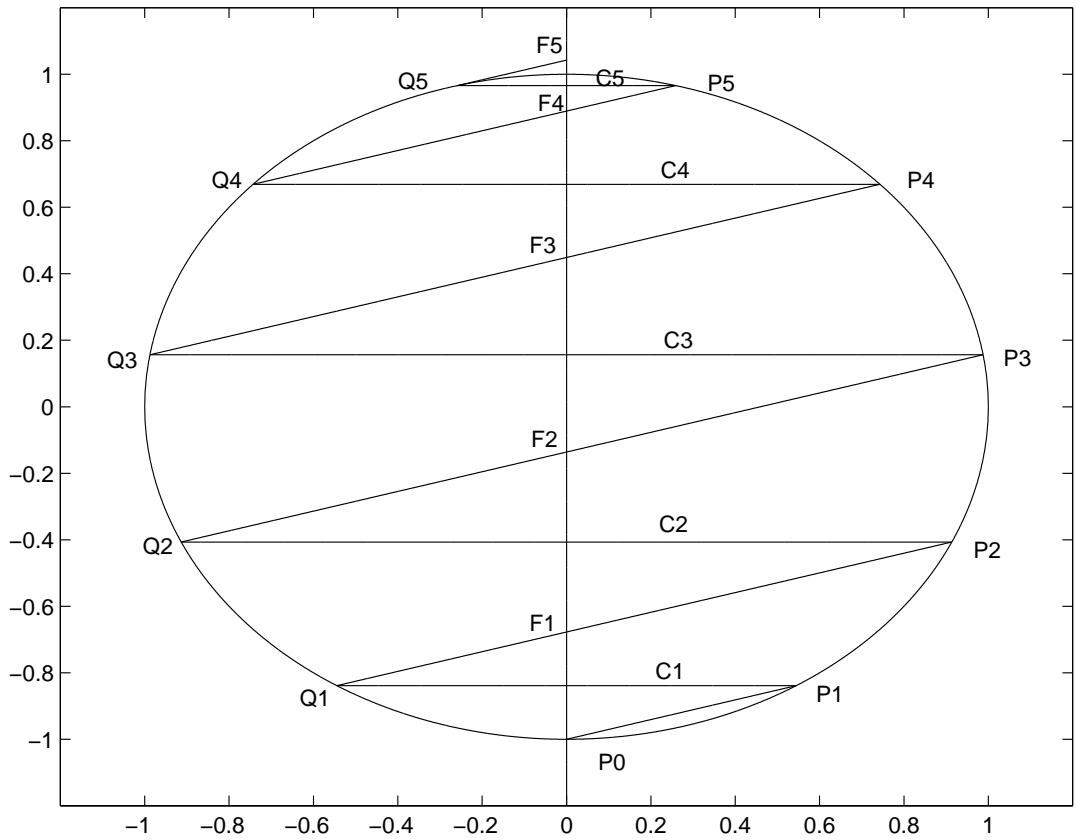}}
\caption{Geometric Construction for Lemmas 3.1 and 3.2}
\end{figure}

\vs

\noindent{\bf Remark:} It is possible to show 
that the sequence $\{z_k\}$ in (\ref{A}) (\ref{B}) 
can be obtained by $z_k=-cos(k \beta)$,  for some
angle $\beta$ obtained as  $\beta=cos^{-1}z_1$. From a geometric point
of view, $\beta$ is the angle in the $y-z$ plane between the segments
${OD_k}$ and ${OD_{k+1}}$, where $O$ denotes the origin and
$D_k:e^{Z_1 \frac{\pi}{2}}P_k$, $k=0,1,2,..., 5$. The points $D_k$ are
the midpoints of the lines representing a circle $C_k$ in Figure 1. This
angle is the same for every $k$. This
gives a geometric interpretation and an alternative proof of Lemma 3.1.  \qed

The above Lemma solves the 
problem of finding the minimum 
number of factors to express 
$X_f$ in the form (\ref{EU73}) 
if we assume that the last term 
on the right is of the type 
$e^{Z_1t}$. This assumption  can be 
relaxed by considering a change 
of coordinates $\tilde T$,
\be{tildeT}
\tilde T:=\pmatrix{\frac{-\rho}
{\sqrt{1+ \rho^2}} & 0 &
\frac{1}{\sqrt{1+ \rho^2}} \cr
0 & 1 & 0 \cr
\frac{1}{\sqrt{1+ \rho^2}} & 
0 & \frac{\rho}{\sqrt{1 + \rho^2}}}.    
\ee    
We have
\be{impo1}
\tilde T Z_2 \tilde T^T=-\sqrt{1+ \rho^2} Z_1, 
\ee
\be{impo2}
\tilde T Z_1 \tilde T^T=-\frac{1}{\sqrt{1 + \rho ^2}} Z_2. 
\ee
Assume that 
\be{lopl}
X_f=e^{Z_1t_1} e^{Z_2 t_{2}} 
\cdot \cdot \cdot e^{Z_2 t_s}, 
\ee
with a term of the type $e^{Z_2 t}$ 
first on the right, is the optimal 
factorization. Then a factorization 
with a term $e^{Z_1 t}$ first on the
right and $s$ factors is the optimal 
factorization for $\tilde T X_f \tilde T^T$ and
viceversa. Therefore according to Lemma 3.2 we have  
$MIN(X_f)={\cal O}(\tilde T X_f \tilde T^T)$. We conclude with the
following Theorem. 

\vs 

\noindent{\bf Theorem 3.3} {\it 
\be{Finalmente}
MIN(X_f)= \min \{ {\cal O}(X_f), {\cal O}(\tilde T X_f \tilde T^T) \}, 
\ee
with $\tilde T$ given in (\ref{tildeT}). 
}

\section{Evaluation of the Generalized Euler Angles} 

The geometric analysis  of the previous section gives a method to
determine the generalized Euler parameters corresponding to the optimal
factorization. Let us assume, without loss of generality, that
$MIN(X_f)={\cal O}(X_f)$ namely, the optimal factorization has a
nontrivial term of the type $e^{Z_1t}$ last on the right. Referring to
Figure 2, we have labeled each region with a number denoting the minimum
number of factors needed to drive $P_s$ to $P_f$ in that region
(Including the last  nontrivial factor on the right  of the type
$e^{Z_1t}$). If $P_f:=-[x_{1,3}, x_{2,3}, x_{3,3}]^T$ is in an odd
region, such as $P_o$ in Figure 2, (namely it is strictly above a plane 
dividing a region between two planes $z=constant$, except for the Region
3, which includes points strictly below the plane $x+ \rho(z+1)=0$ as
well) then an optimal factorization for $X_f$ is 
\be{FACT1}
X_f=e^{Z_1 t_1} e^{Z_2 t_2} e^{Z_1 \pi} e^{Z_2 \pi} 
\cdot \cdot \cdot e^{Z_1 \pi} e^{Z_2 \pi} e^{Z_1 t_s}. 
\ee     
We first determine $t_2$ so that, defined $L:=e^{Z_2 t_2} e^{Z_1 \pi} e^{Z_2 \pi} 
\cdot \cdot \cdot e^{Z_1 \pi} e^{Z_2 \pi}:= \{ l_{i,j} \}$,
$l_{3,3}=x_{3,3}$. Then we determine $t_1$ so that $e^{Z_1 t_1}L
P_s=P_f$, where $P_s$ denotes the South Pole $P_s=[0,0,-1]^T$ 
and then $t_s$ such that $e^{Z_1 t_1} L e^{Z_1
t_s}=X_f$. Notice that each step involves the evaluation of just one
parameter. Notice also that the optimal factorization is not unique and,
in the above factorization, we could have, for example,  replaced the
term  $e^{Z_2 t_2} e^{Z_1 \pi}$ with a term $e^{Z_2 \bar t_2} e^{Z_1
\bar t_1}$ for appropriate values $\bar t_1$ and $\bar t_2$ (see the
alternative path with bold face lines in Figure 2). If 
$MIN(X_f)$ is even ($P_f=P_e$ in
Figure 2) then we have that the optimal factorization is given by 
\be{FACT2}
X_f=e^{Z_2 t_1} e^{Z_1 t_2} e^{Z_2 \pi} e^{Z_1 \pi} 
\cdot \cdot \cdot e^{Z_1 \pi} e^{Z_2 \pi} e^{Z_1 t_s}. 
\ee
Let, in the sequence (\ref{A}), (\ref{B}), $\bar z_k$ be the largest value of
$z_k$ such that $z_k < -x_{3,3}$. Then, we consider a point $\bar
P:=[\bar x, \bar y, \bar z]^T$ intersection of the planes 
$(x+x_{1,3})+ \rho (z+ x_{3,3})=0$, $z=\bar z_{k}$ and the sphere 
$x^2+y^2+z^2=1$. Then we determine $t_2$ so that, defined  
$L:= e^{Z_1 t_2} e^{Z_2 \pi} e^{Z_1 \pi} \cdot \cdot \cdot 
e^{Z_1 \pi} e^{Z_2 \pi} P_s$, we have $ L P_s = \bar P$. Then we
determine $t_1$ so that $e^{Z_2 t_1} \bar P=P_f$ and finally we
determine $t_s$ so that $e^{Z_2 t_1}L e^{Z_1t_s}=X_f$. In this case too,
the optimal factorization is not unique.

\begin{figure}
\epsfysize=6.50in
\epsfysize=4.50in
\centerline{\epsfbox{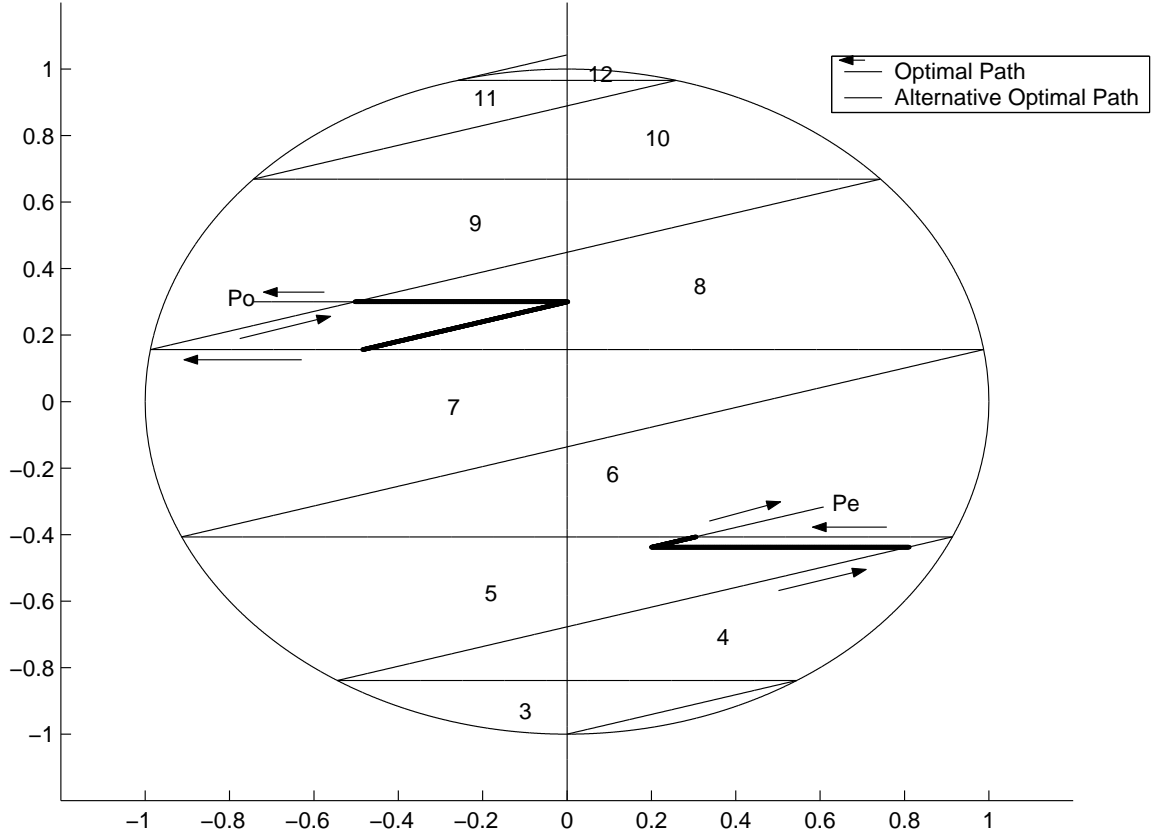}}
\caption{ Optimal paths on the sphere}
\end{figure}

\section{Applications} 

The results of this paper can be used to prescribe bang bang type of 
controls for bilinear systems whose state varies on the Lie group
$SO(3)$,  with minimum number of switches. 
This technique of control can be applied to the dynamics of a
rigid body where the angular velocity is  seen as control. The same
technique can  also
be employed for  the control of switched electrical networks 
\cite{RamakCDC}, with minimum number of switches.

In recent years there has been a large amount of interest in the control
of systems of the form (\ref{EU71}) with $A$ and $B$ in the Lie algebra
$su(2)$. This Lie algebra is 
isomorphic to $so(3)$. These systems model the dynamics of two
level quantum systems with just one control \cite{conmohIEEE}. Constructive 
factorizations (\ref{EU73}) 
of elements of the Lie group $SU(2)$ have been given 
in \cite{HOMO}, \cite{RAMA1},  and used for control. In particular the
 factorization of \cite{HOMO} gives a worst case number of factors
which is greater than the minimum by at most one. 
The algorithm presented in this paper can be used to determine the
optimal factorization for elements $\bar X_f$ in $SU(2)$ as well and
therefore to prescribe a control algorithm for two level quantum systems
with minimum number of switches. 

\vs

Let 
$\tilde \phi$ denote the isomorphism between $su(2)$ and $so(3)$ which
maps the Pauli matrices 
\be{BeatiPauli}
S_x:=\pmatrix{0 & -i \cr 
-i & 0}, \quad S_y:=\pmatrix{0 & -1 \cr 1 & 0} 
\quad 
S_z:=\pmatrix{-i & 0 \cr 0 & i},  
\ee    
to $2S_{1,3}$, $2S_{2,3}$, $-2 S_{1,2}$ respectively. Let $\bar Z_1$ and
$\bar Z_2$ be two linearly independent matrices in $su(2)$. We
look for the factorization of $\bar X_f$ of the type 
\be{P1}
\bar X_f=e^{\bar Z_1 t_1} e^{\bar Z_2 t_2} \cdot \cdot \cdot 
e^{\bar Z_s t_s},   
\ee
with minimum number of factors. The isomorphism $\tilde \phi$ between 
$su(2)$ and $so(3)$ induces a homomorphism $\phi$ between elements of the
corresponding Lie groups, $\phi: SU(2) \rightarrow SO(3)$, which is
given, if $S =e^{V} \in SU(2)$, by $\phi(S):=e^{\tilde \phi(V)}$. This
homomorphism is two to one in that to $\pm S$ in $SU(2)$ 
corresponds   the 
same element in $SO(3)$ (for a more detailed treatment of the relation
between the Lie groups $SU(2)$ and $SO(3)$ 
see e.g. \cite{sternberg}. See also \cite{ioSCL} for applications to
control). Let $X_f$ be the element in $SO(3)$ corresponding to $\bar X_f$
under this homomorphism and $Z_1$ and $Z_2$ the elements of $so(3)$
corresponding to $\bar Z_1$ and $\bar Z_2$. If 
\be{ancora}
X_f= e^{Z_1 t_1} e^{Z_2 t_2} \cdot \cdot \cdot 
e^{Z_s t_s}, 
\ee
is the optimal factorization for $X_f$ then $s$ is the optimal number of
factors for $\bar X_f$ in (\ref{P1}) as well. The generalized 
Euler parameters can also be easily determined. If we use the same values for
$t_1,...,t_s$ in (\ref{ancora}) and  (\ref{P1}) we obtain a matrix which
is $\pm$ the desired  $\bar X_f$. This affects the quantum mechanical 
state $x$ in 
(\ref{EU71}) by an overall  phase factor which has no physical meaning. 
In any case,  the minus sign  can be easily eliminated by
changing the value of just one of the parameters so as to change one
factor $S$ into $-S$. This is always possible since each 
one dimensional subgroup in $SU(2)$ that contains $S$ also contains
$-S$. Therefore, we can find an optimal factorization for any element in
$SU(2)$ as well. This can be easily extended to any Lie algebra
isomorphic to $su(2)$ and the 
corresponding Lie group, which is known to be isomorphic to either
$SO(3)$ or $SU(2)$ \cite{KL1}.

%%%%%%%%%%%%%%%%%%%%%%%%%%%%%%%%%%%%%%%%%%%%%%%%%%%%%%%%%%%
%%%%%%%%%%%%%%%%%%%%%%%%%%%%%%%%%%%%%%%%%%%%%%%%%%%%%%%%%%%%
%%%%%%%%%%%%%%%%%%%%%%%%%%%%%%%%%%%%%%%%%%%%%%%%%%%%%%%%%%%%
%%%%%%%%%%%%%%%
%%%%%%%%%%%%%%%
%%%%%%%%%%%%%%%
%%%%%%%%%%%%%%%
%%%%%%%%%%%%%%%
%%%%%%%%%%%%%%%%%%%%%%%%%%%%%%%%%%%%%%%%%%%%%%%%%%%%%%%%%%%%
%%%%%%%%%%%%%%%%%%%%%%%%%%%%%%%%%%%%%%%%%%%%%%%%%%%%%%%%%%%%
%%%%%%%%%%%%%%%%%%%%%%%%%%%%%%%%%%%%%%%%%%%%%%%%%%%%%%%%%%%%

\end{document}